\newcommand{\f}[1]{Fig.~\ref{#1}}
\newcommand{\eq}[1]{Eq.~(\ref{#1})}

\def\d{\partial}
\def\be{\begin{equation}}
\def\ee{\end{equation}}
\def\bea{\begin{eqnarray}}
\def\eea{\end{eqnarray}}
\def\l({\left(}
\def\r){\right)}

\def\Bfj{B_{\rm fj}}

\documentstyle[prb,aps,multicol,graphicx,psfig]{revtex}

 \renewcommand{\narrowtext}{\begin{multicols}{2} 
\global\columnwidth20.5pc}
 \renewcommand{\widetext}{\end{multicols} \global\columnwidth42.5pc}

\begin{document}
\title{Local threshold field for dendritic instability in 
superconducting MgB$_2$ films}
\author{F. L. Barkov$^{1,2}$, D.~V. Shantsev$^{1,3}$, T.~H.~Johansen$^{1,}$\cite{0},  
P. E. Goa$^1$,  \\
W. N. Kang$^{4}$, H. J. Kim$^{4}$, E. M. Choi$^{4}$ and 
S. I. Lee$^{4}$ 
}
\address{
$^1$Department of Physics, University of Oslo, P. O. Box 1048
Blindern, 0316 Oslo, Norway\\
$^2$Institute of Solid State Physics, Chernogolovka, Moscow Region, 142432, Russia\\
$^3$A. F. Ioffe Physico-Technical Institute, Polytekhnicheskaya 26,
St.Petersburg 194021, Russia\\
$^4$National Creative Research Initiative Center for Superconductivity, 
Department of Physics, Pohang University of 
Science and Technology, Pohang 790-784, Republic of Korea
}
%\date{\today}
\maketitle

\vspace{-4.8cm}
\centerline{Submitted to Phys. Rev. B on 17.05.2002, cond-mat/0205361}
\vspace{4.8cm}

\begin{abstract}
Using magneto-optical imaging the phenomenon of dendritic flux penetration 
in superconducting films was studied. Flux dendrites were abruptly formed in a
300~nm thick film of MgB$_2$ by applying a perpendicular magnetic field.
Detailed measurements of flux density distributions show that there exists
a local threshold field controlling the nucleation and termination 
of the dendritic growth. At 4 K the local threshold field
is close to 12 mT in this sample, where the critical current density is
$10^7$~A/cm$^2$. The dendritic instability in thin films is believed to be 
of thermo-magnetic origin, but the existence of a {\em local} threshold field, 
and its small value are features that distinctly contrast the thermo-magnetic 
instability (flux jumps) in bulk superconductors.
\end{abstract}


\narrowtext
~

\vspace{-1.8cm}
\section{Introduction}

An abrupt penetration of magnetic flux in the form of
 branching patterns was
first observed  in 1967, in superconducting Nb alloys.\cite{1967}
The phenomenon received much attention in the 1990s, when 
advancements in magneto-optical (MO) imaging allowed studies
with a much higher spatial resolution.
The branching phenomenon, or dendritic instability, has now been observed
in YBa$_2$Cu$_3$O$_7$ films\cite{leiderer,bolz} (induced by a laser pulse), 
in field-cooled Nb films\cite{duran} and in zero-field-cooled (ZFC) 
patterned Nb films.\cite{vv}
However, the material most sensitive to the instability seems to be
the recently discovered superconductor MgB$_2$, the only material
where the flux dendrites appear in uniform ZFC films placed in an 
applied field or triggered by passing a transport current.\cite{epl,sust,phc,apl}

The dendritic flux instability is believed to be of thermo-magnetic origin, 
similarly to the much more explored phenomenon of flux jumps.
Flux jumps is a dominant threat to the stability of
the critical state in superconductors, and is especially
important in high-current and
high-field applications.\cite{swartz,wipf-prb,wipf,mints}  
Local heating due to motion of the magnetic vortices 
reduces the pinning, and will facilitate their further motion.
This may lead to an avalanche process, where a macroscopic
amount of flux suddenly invades the superconductor -- a process accompanied
by a strong heating.

A number of common features indicate that the
same physical mechanism is underlying both
the dendritic instability and flux jumps.
First, both phenomena occur only at low temperatures, and they
develop very fast ($10^4-10^6$~cm/s, see 
Refs.~\onlinecite{1967}, and~\onlinecite{leiderer}). Moreover, 
both instabilities can be suppressed by contacting the superconductor 
with normal metal so that heat is removed more efficiently\cite{phc,harr2}
Furthermore, dynamics in the form of branching flux and temperature distributions 
has recently been obtained by 
computer simulations accounting for the heat produced by flux 
motion,\cite{epl,aranson} and thus support strongly that the flux dendrites
indeed results from a thermo-magnetic instability.  

There seems to be two necessary conditions for
an instability to develop along the dendritic scenario. 
The first is a small thickness of the superconductor: the dendrites
have so far been observed only in films 
with thickness $\le 0.5~\mu$m.\cite{thick1967}
The long-range vortex-vortex interactions typical for thin 
films\cite{pearl,brandt-rev}
are probably essential for the branching flux structures to form.
Secondly, the process should be adiabatic so that 
the temperature distribution remains highly nonuniform during
the dendrite growth. Computer simulations\cite{aranson} have
demonstrated that dendrites occur only when the  
heat diffusivity is much smaller than the magnetic diffusivity.

The key quantity characterizing flux-jump is the 
applied field, $\Bfj$, when the first jump occurs in a ZFC superconductor.
The $\Bfj$ also determines the interval between complete flux jumps
(the magnetization dropping to zero) as seen in \f{f_2types}(left).
Therefore, the central  question is: {\em ``Does a threshold field exist 
also for dendritic flux jumps?''}.%
\begin{figure}[p]
\centerline{\includegraphics[width=8cm]{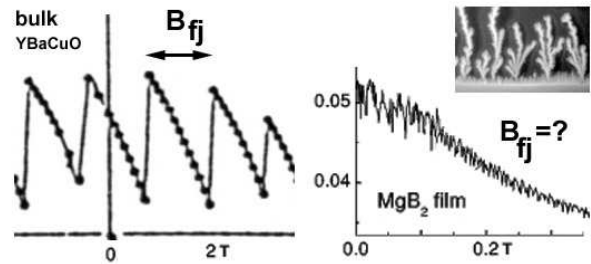}}
\vspace{0.1cm}
\caption{Magnetization $M(B)$ data exhibiting 
conventional flux jumps (left, Ref.\protect\onlinecite{chen}),
and irregular small jumps (right, Ref.\protect\onlinecite{zhao})
due to abrupt penetration of flux dendrites shown in the inserted MO image. 
%Determination of the threshold field $B_{fj}$ for dendritic flux jumps
%is the subject of the present study.   
\label{f_2types}}
\end{figure}
{}From typical $M(B)$ data for an MgB$_2$ film shown in \f{f_2types}(right) 
the answer seems ``No''.  The jumps are here very small, 
typically $\Delta M \sim 10^{-2}M$,
because one dendritic structure occupies only a small fraction
of the sample. Furthermore, they are irregularly spaced along the 
applied field axis, and the exact jump pattern is irreproducible 
when the experiment is repeated.

In the search for a ``dendritic'' $\Bfj$ we have performed a
magneto-optical study of flux penetration in a virgin MgB$_2$ film, and analyzed
quantitatively the flux density distributions produced by the dendritic instability.
We find that the dendritic instability indeed has a threshold field.
However, this is a threshold not for the applied field but instead for
the local flux density. 
This {\it local} threshold field determines when and where
in the superconducting film the dendritic structures nucleate.

The paper is organized as follows. In section II the sample and the experimental 
method used in this work are briefly described. The results of the MO imaging 
investigation are presented in section III, and a discussion follows in
section IV. Finally, the conclusion are given in section V.

\section{Experimental}

Films of MgB$_2$ were fabricated on Al$_2$O$_3$
substrates using pulsed laser deposition.\cite{kang}
A 300~nm thick film shaped as a square 
with dimensions 5$\times$5~mm$^2$ was selected for the present studies.
The sample has a high degree of $c$-axis alignment perpendicular to the plane,
and shows a sharp superconducting transition at $T_c=39$~K.

The flux density distribution in the superconducting film was visualized
using MO imaging based on the Faraday effect in ferrite garnet indicator films. 
For a recent review of the method, see Ref.~\onlinecite{jooss}, and a description of our 
setup is found elsewhere.\cite{joh96}
The sample was glued with GE varnish to the cold finger of the optical cryostat, and
a piece of MO indicator covering the sample area was placed loosely on top of 
the MgB$_2$ film. Before the mounting
a few plastic spheres of diameter 3.5~$\mu$m was distributed over
the sample surface to avoid thermal influence of the MO indicator.

As usual, the gray levels in the MO images were converted to magnetic field values
using a calibration curve obtained above $T_c$. In all the images shown in the 
present work, the bright regions correspond to high values of the
flux density, while the fully dark areas are free of flux, i.e. Meissner state regions.   
All the experiments were carried out at $T=3.6$~K on an initially ZFC sample.

\section{Results}

%\subsection{Images}

The MgB$_2$ film was placed in a slowly increasing perpendicular 
applied field, $B_a$.
At small fields, up to $B_a=2$~mT, we observed just conventional 
flux penetration where a 
gradual increase in $B_a$ results in a smooth advancement of the flux front.
%For $B_a=2$~mT the flux penetrated 20~$\mu$m into the film.
Increasing the field further this smooth behavior starts to be 
accompanied by a sudden invasion of macroscopic dendritic structures, 
as illustrated in \f{f_images} showing MO images of the
flux distribution near the edge at $B_a= 2.3$, 3.2 and 7.4~mT.
The images cover different parts of the sample which have in common that
dendritic structures had been formed just before the images were recorded.
As in earlier studies, these structures are seen to
develop at seemingly random places that vary from
one experiment to another, and the dendrites grow to final size 
faster than we can detect (1~ms).   
As the applied field continues to increase we find that
outside the dendritic areas the flux front advances gradually,
while the dendrites that are already formed always remain completely frozen.
Below we focus on the dendritic flux behavior, whereas
the gradual "background" penetration, also displaying interesting
non-uniform features, will be the subject of a separate paper. 

\widetext
\begin{figure}
\centerline{
\includegraphics[height=5cm]{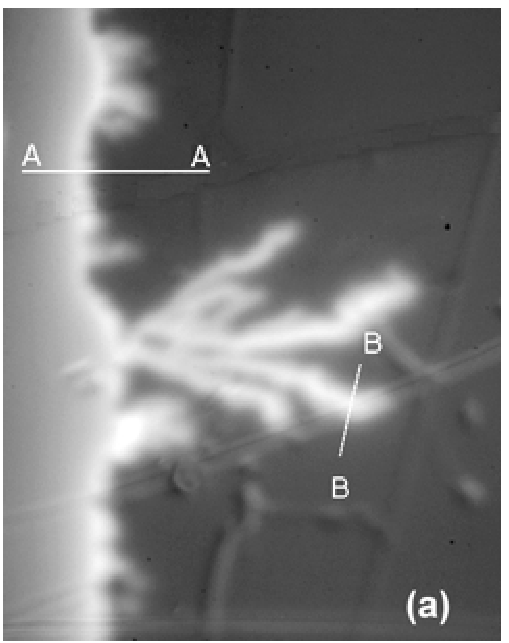}
\includegraphics[height=5cm]{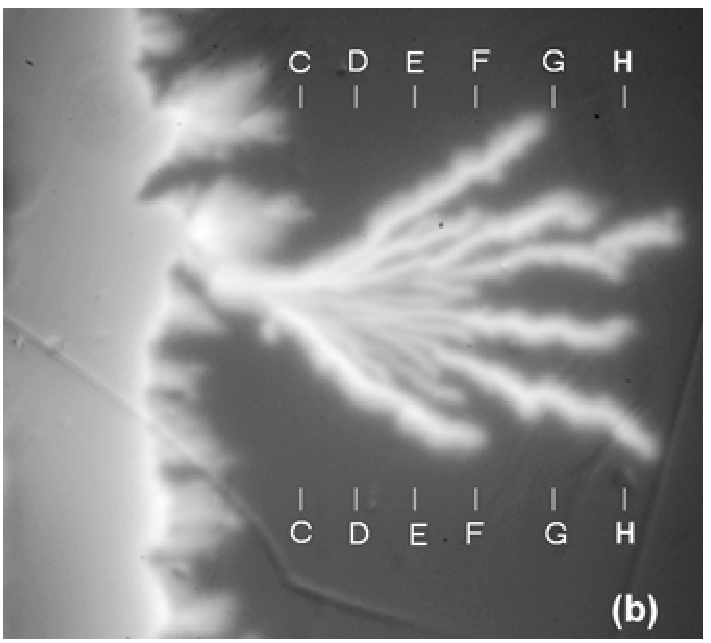}
\includegraphics[height=5cm]{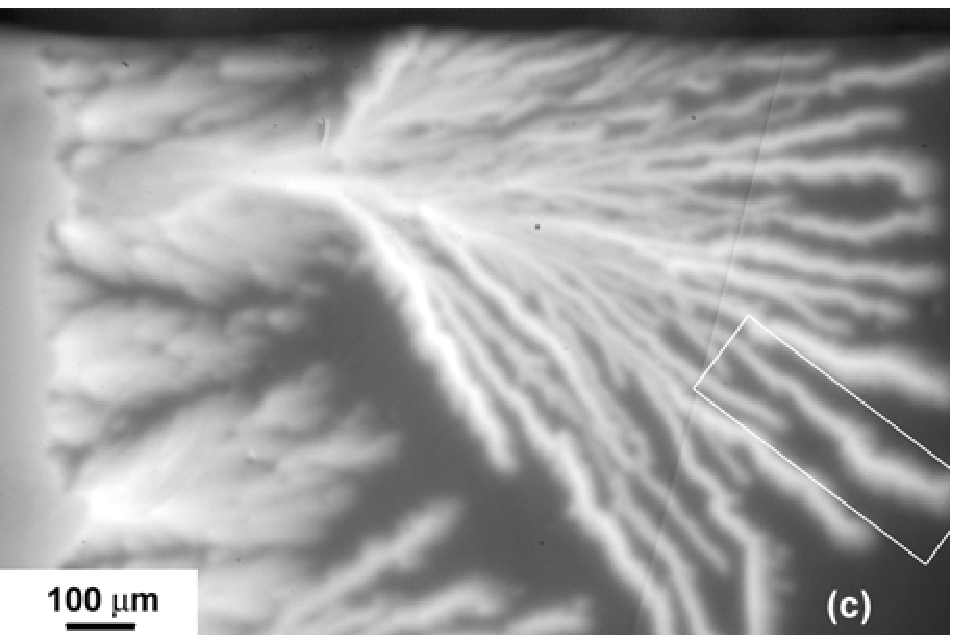}
}
\vspace{0.6cm}
\caption{MO images showing
dendritic flux structures formed near 
the edge of the MgB$_2$ film at applied fields, which in (a)-(c)
are $B_a=2.3$, 3.2 and 7.4~mT, respectively. 
The dendritic structures for different $B_a$ differ in size, 
but not in the flux density (image brightness) along the core of the
individual branches.   
\label{f_images}}
\end{figure}
\narrowtext

%\subsection{One branch}

\begin{figure}
\centerline{\includegraphics[height=6.5cm]{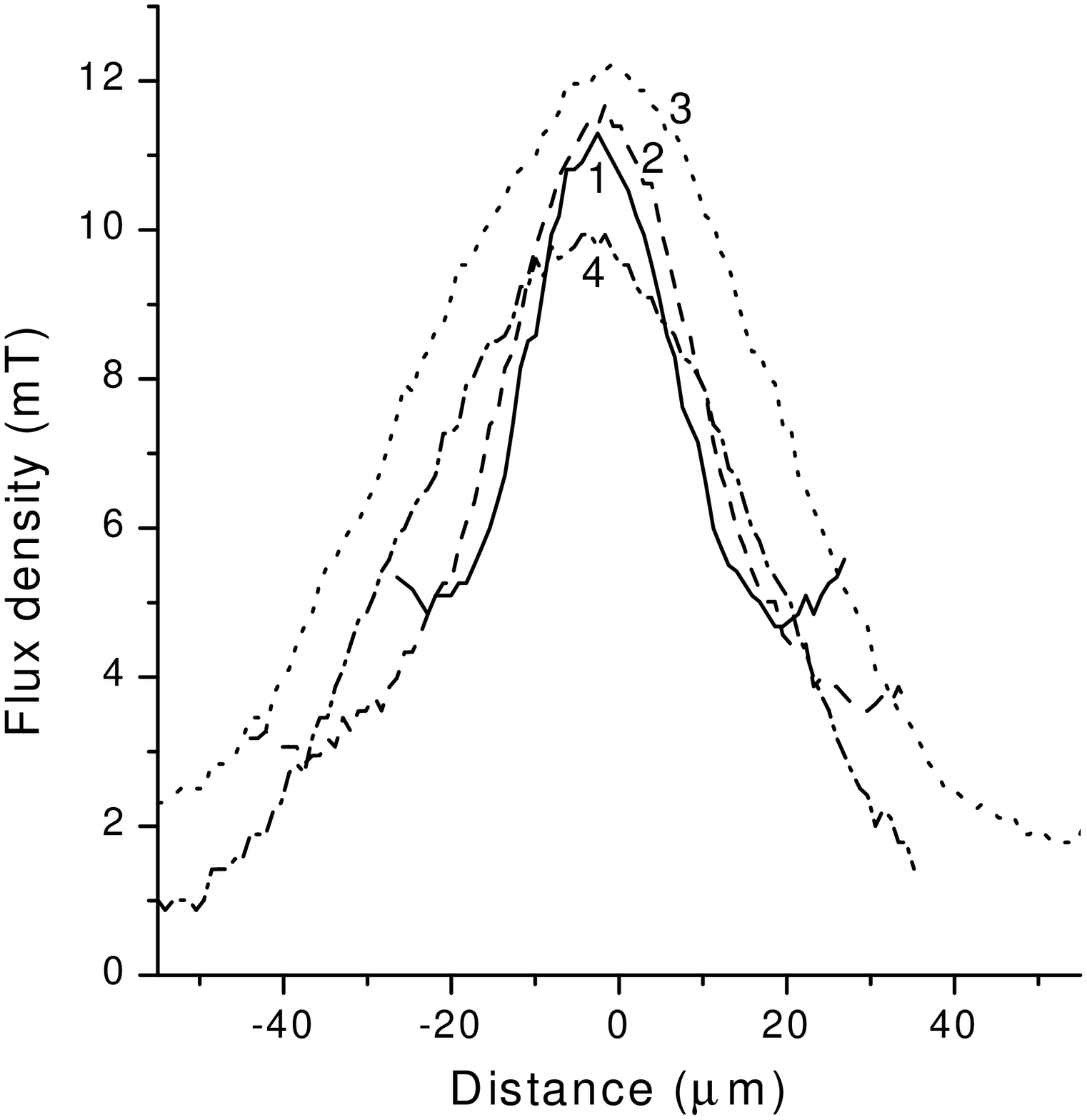}
\includegraphics[height=6.5cm]{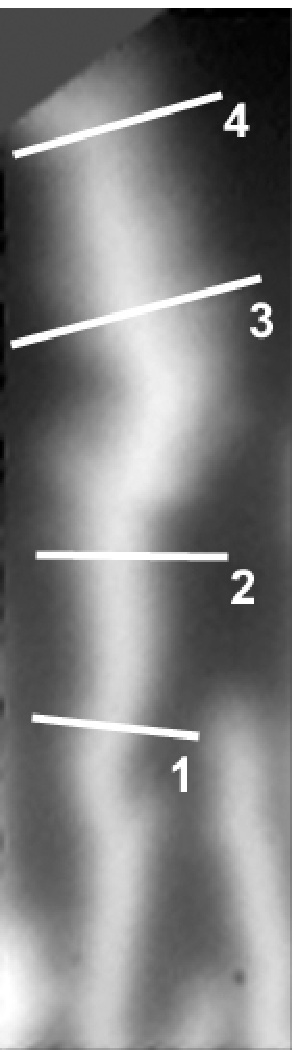}}
\vspace{0.3cm}
\caption{Profiles of flux density across one branch of a large 
dendritic structure which appeared at $B_a=7.4$~mT.
The MO image shows the region marked by the rectangle in \f{f_images}(c). 
\label{f_finger}}
\end{figure}

 \begin{figure}
\centerline{\includegraphics[width=8.2cm]{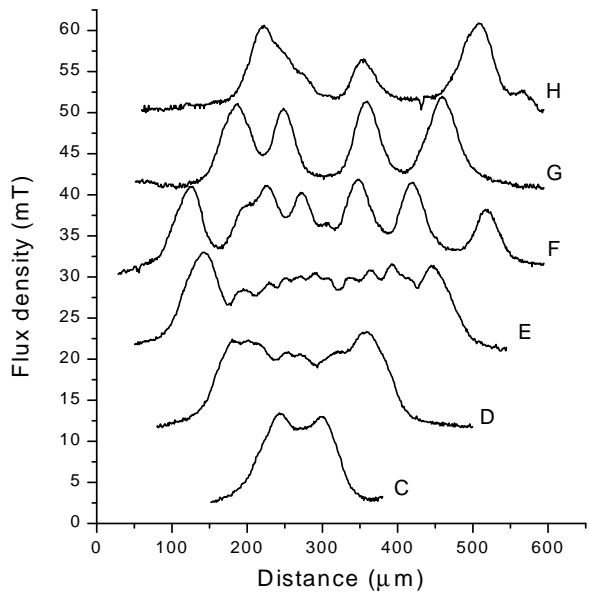}}
\vspace{0.3cm}
\caption{Flux density profiles across a  
dendritic structure that appeared at $B_a=3.2$~mT, and is
shown in \f{f_images}(b). To avoid overlap
the profiles are shifted (by 10~mT) with respect to each other.  
\label{f_tree}}
\end{figure}

Let us first look more detailed on one individual branch of a dendritic 
structure, where the area marked by a white rectangle in \f{f_images}(c)
was chosen as a typical example. Several flux density profiles 
across the large branch were measured with the result seen in \f{f_finger}.
The profiles have an overall triangular shape that 
varies only little along the branch.
The maximum flux density in the center remains essentially the same,
10-12~mT, whereas %except near the very end of the branch.
the branch width increases from $30~\mu$m
near the root towards $ 50~\mu$m near the tip.
The reason for such broadening is not fully clear, but is probably 
related to the fact that near the tip the distance from  
neighboring branches is larger.
Near the root of a dendritic structure the branches always grow densely
and their mutual repulsion causes each of them to be compressed.

Shown in Fig.~\ref{f_tree} is a series of flux density profiles across 
the whole dendritic ``tree" seen in \f{f_images}(b).
To avoid overlap of graphs, 
the subsequent profiles C\ldots H are shifted along the vertical axis.
By comparing the two figures each peak can be identified as a branch present
in the MO image. For the profiles C-E the outmost branches of the tree
are the more pronounced, and many minor ones are located in between.
In the profiles F-H the number of branches reduces, and large flux-free
regions exist between them. From this set of curves we see that, like in
\f{f_finger}, the maximum flux density (at the dendrite core) remains
essentially constant, $B_{\max}\approx 12$~mT, along one branch. 
Moreover, a striking fact is that this value is the same for all branches.
% and is even 
%close to the peak value of $B$ at the film edge, as is seen from
%the profile A, plotted as a dashed curve in Fig.~\ref{f_tree}.
%\subsection{Whole tree}
%\subsection{Edge \& dendrite}

This universality motivates a comparison also with the $B$ profile 
across the film edge.
Due to demagnetization effects the field becomes concentrated
near the edge of a superconducting film, and hence the MO images 
show a line of maximum brightness exactly along the sample edge. 
For the comparison we choose the MO image from \f{f_images}(a), 
where the flux penetration near the edge is most regular. 
The flux density profiles along the lines A-A and B-B are presented 
in \f{f_2prof}.
Note that in the A profile only the part to the right of the peak 
is relevant, i.e., representing penetrated magnetic flux.
Remarkably, we find that the peak values as well as the slopes of the 
profiles are essentially the same. 
%The same holds true for profiles measured at other applied fields.
Hence, this together with similar investigations 
made at other locations lead us to conclude that the MgB$_2$ film does not 
allow anywhere in the sample the 
local field to exceed the value of $B_{\max}\approx 12$~mT.  

\begin{figure}
\centerline{\includegraphics[width=8.2cm]{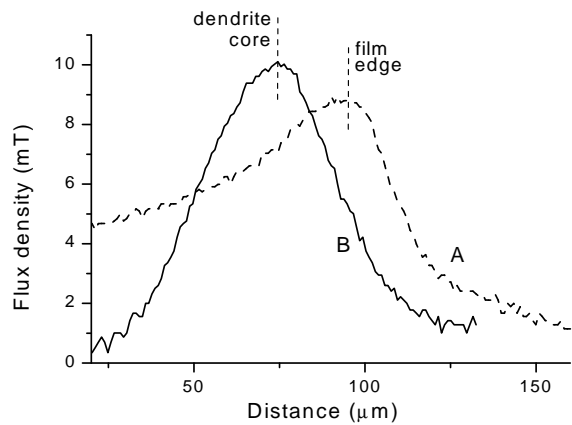}}
\vspace{0.3cm}
\caption{Profiles of flux density  
near the edge (A), and across a dendrite (B)
obtained from the MO image shown in \f{f_images}(a). %$B_a=2.3$~mT.
\label{f_2prof}}
\end{figure}

How this universality of the $B_{\max}$ applies very generally
can be illustrated by histograms of the flux density distributions
at different applied field.
Shown in \f{f_hist} are histograms of $B(x,y)$ over the entire field-of-view
of the MO images for three applied fields in the range of 2-8~mT. 
At 2.3~mT the number of dendrites is small and so is their size, and only a 
very small fraction of the sample are has a local field exceeding 10~mT.
As the applied field increases the histogram develops a pronounced peak near
10~mT and the existence of a maximum local field becomes evident. 
It is clear that
although the dendrites that are formed vary widely in their size, and the total
area covered by dendrites is very different,
the maximum local field remains the same. 

\begin{figure}
\centerline{\includegraphics[width=8.2cm]{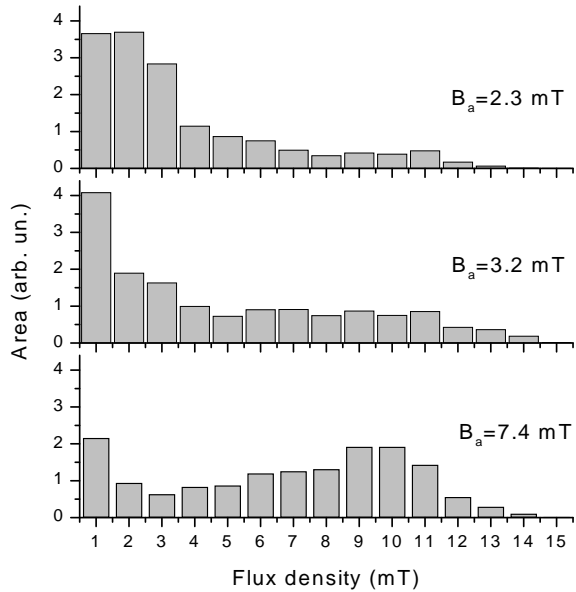}}
\vspace{0.3cm}
\caption{Histograms of areas having various flux densities
(pixel values of the MO images in \f{f_images}) in the sample 
at 3 applied fields. Despite the quite different $B_a$ 
all histograms display approximately the same maximum field of 12~mT.     
\label{f_hist}}
\end{figure}

\section{Discussion}

The existence of the universal field $B_{\max}$ sheds new light on the
mechanism of dendrite formation. 
The present results suggest that $B_{\max}$ serves as a {\em local threshold} 
 field for nucleation of the instability.
When the ZFC superconducting film is placed in an increasing applied field, 
its interior remains first flux free, and the field becomes enhanced
at the edges. When the local edge field exceeds $B_{\max}$,
an avalanche-like invasion of flux is nucleated there.
The avalanche development is then driven not only by local heating, 
as in bulk superconductors,
but in a film also by a local increase of the field.
Geometrical field amplification at concave defects along the 
edge is an effect well-known from MO imaging studies,\cite{jooss,kobl,defect}
and is due to the bending of the Meissner currents that are forced to 
flow around the defected region, in this case the heated spot where the
avalanche starts. In the first stage the avalanche grows
predominantly in the direction perpendicular to the edge, as seen from the 
distinct root of the dendritic structures. 
In a similar way, any perturbation along this penetration channel will be  
enhanced due to the Meissner current bending, and eventually result in 
multiple branching. Note that the irreproducibility of the dendritic 
patterns indicate that the branching points are not directly related to 
the pinning landscape or other non-uniformities grown into the sample.
%A necessary condition for the invasion to end is that the flux density is 
%smaller than $B_{\max}$ everywhere in the film.
The dendritic structure will grow and continue branching until the 
flux density reduces to $B_{\max}$ in the cores of all branches.
This is the final state of the instability, and is what we see in the 
MO images.
Upon further increase of the applied field, the condition $B<B_{\max}$
becomes soon violated again, now at a different place along the edge from
where a new dendritic structure will invade the film.

Another interesting result most clearly seen from \f{f_2prof}
is that the slope of the $B$ profiles near the film edge,
and across a dendritic branch is essentially the same.
Actually, this slope is also characteristic for all profiles across the
large dendritic structure shown in \f{f_tree}. Since the slopes
reflect the local critical current density, $j_c$,
this suggests that they were formed
at the same temperature, since $j_c$ is strongly temperature dependent.
It is then clear from \f{f_2prof} that 
the heating during the dendrite propagation stage was localized 
to a very narrow core of the branch. The sharpness of the peaked
flux profiles shows that the heated core has a width of 15~$\mu$m, 
or less. 
%while the side regions remained at the initial temperature.
This is consistent with results obtained also by Bolz 
et al.\cite{boltz}

%Ref.\onlinecite{leiderer},
%where magneto-optical imaging with nanosecond resolution
%was used to study dendritic instability in YBa$_2$Cu$_3$O$_7$ films.
%It was found that at first flux penetrates along the cores
%of the dendrites, and only later the slopes of the channels are formed. 

The magnitude of the critical current density can be estimated
from the profiles of \f{f_2prof} using the Bean model
formula for a thin strip in a perpendicular field.\cite{BrIn,zeld}
For small fields, the flux penetration depth is given as 
$\delta = 0.5 w (\pi B_a/\mu_0 d j_c)^2$, where $w$ is the strip halfwidth,
and $d$ is its thickness.
Substituting here $\delta=30~\mu$m, and $w=2500~\mu$m one obtains
 $j_c \approx 10^7$~A/cm$^2$.
Unfortunately, this high $j_c$ is compromised by the dendritic instability 
which largely influence the macroscopic magnetic properties of the film.
Magnetization curves of MgB$_2$ films in general (i) contain numerous
dips of different magnitudes (noisy behavior) and
(ii) the averaged $M$ recalculated into critical current density
gives a much lower value than the "true" $j_c$.\cite{zhao,epl}

Let us compare the experimentally found threshold field 12~mT with
theoretical estimates. The first jump field within the adiabatic
approach and the Bean model is given as\cite{swartz,wipf-prb,wipf,mints}
\be
\Bfj = \left(\frac{2\mu_0 C j_c}{\d j_c/\d T} \right)^{1/2},
\label{bfj} 
\ee
where $C$ is the heat capacity.
This result is obtained for bulk superconductors and should be
modified for thin samples in a perpendicular field.
Using the Bean-model flux distributions in thin films\cite{BrIn,zeld}
one can show\cite{ad-strip} that the field $\Bfj$
should be multiplied by a factor $\sim \sqrt{d/w}$.
Then, substituting $C = 0.3$~kJ/Km$^3$  (Ref.~\onlinecite{walti})
% Cp=0.005 J/(mole K), mole = 3 gat, Vgat=5.83 cm^3/gat (cond-mat/0106394)
in \eq{bfj}, assuming $j_c(T) \propto  (T_c-T)$,  
%and mulitplying the result by $\sqrt{d/w}=0.013$ 
we obtain  $\Bfj = 1.5$~mT  at 4~K. 
Note that this $\Bfj$ should give the {\em applied} field
when the first flux dendrite enters a ZFC film.
For our experiments on the MgB$_2$ film this field equals 2~mT, 
in excellent agreement with the theoretical estimate.

At this field, $B_a=2$~mT, the local flux density was measured to be
12~mT (\f{f_2prof}) at the film edge.\cite{why12} 
After many dendrites have entered the film, 
the flux distribution became strongly non-uniform, and 
the criterion for the 
threshold {\em applied} field $\Bfj$ is no longer applicable.
Nevertheless, we find that the value 12~mT can still be 
used as a {\em local} threshold field. 

%This scenario has to be modified when the applied field becomes
%higher than $B_{\max}$.
%Then, the film interior is almost completely covered by dendrites,
%as found in our previous studies.\cite{epl}
%When a new dendrite enters the film, the flux density $B$ in its core 
%is larger than average flux density $\bar{B}$ in the area.
%Probably, the concept of threshold field 
%survives here too, only the condition for nucleation of the instability 
%should be rewritten as $B-\bar{B} > B_{\max}$.
%A quantitative check of this condition seems
%impossible since the flux distribution is severely non-uniform, and
%$\bar{B}$ is very vaguely defined.

\section{Conclusions}

In summary, we have found using magneto-optical imaging,
that the nucleation as well as the termination of
dendritic flux avalanches in superconducting MgB$_2$ films is governed
by a local threshold field, $B_{\max} \approx 12$~mT. Avalanches nucleate 
near the edge whenever the local flux density exceeded $B_{\max}$ and the 
flux invasion proceeds as long as there exist regions where 
$B > B_{\max}$. As a result, each avalanche ends with having a flux distribution
where $B = B_{\max}$ in the cores of all branches of the 
dendritic structure. We find that as the applied field is increased, 
the flux density at the edge remains equal to $B_{\max}$, while
new dendritic flux structures will invade the film
The width of the dendrite cores, which are the heated
channels of flux invasion was found to be 15~$\mu$m, or less, 
whereas in the final state the dendrite fingers are 60-80~$\mu$m wide.

\section*{Acknowledgements}
 
The financial support from the Research Council 
of Norway, the Russian Foundation for Basic Research (grant 01-02-06482),
and the Ministry of Science and Technology of 
Korea through the Creative Research Initiative Program 
is gratefully acknowledged.

\widetext

\begin{references}
\bibitem{why12} The measured value 12~mT 
is sensitive to the distance between the MO indicator and the 
superconducting film,
which was $\sim$5~$\mu$m in the current setup.
Measurements directly at the surface would give a slightly larger value,
but clearly not change the fact that a local threshold field exists.
\end{references}
\end{document}